\documentclass[]{spie}  

\usepackage[]{graphicx}
\usepackage{svg}
\usepackage{multirow}
\usepackage[]{amsmath}
\usepackage{fancyhdr}

\title{Learning from our neighbours: a novel approach on sinogram completion using bin-sharing and deep learning to reconstruct high quality 4DCBCT} 

\author{Joel Beaudry\supit{a}\supit{b}\supit{$\dagger$} and Pedro L. Esquinas\supit{c}\supit{$\dagger$}
\skiplinehalf
\supit{a}Department of Radiation Oncology (University of Toronto), Toronto, Canada; \skiplinehalf
\supit{b}Durham Regional Cancer Centre (Lakeridge Health), Oshawa, Canada; \skiplinehalf
\supit{c}Department of Radiology (University of British Columbia), Vancouver, Canada
}

\authorinfo{Author correspondence: \\Joel Beaudry: jbeaudry@lakeridgehealth.on.ca\\  Pedro L. Esquinas: esquinas@mail.ubc.ca \\
\supit{$\dagger$} Contributed equally}

\begin{document} 
  \maketitle 
\thispagestyle{fancy}

\begin{abstract}
Inspired by the success of deep learning applications on restoration of low-dose and sparse CT images, we propose a novel method to reconstruct high-quality 4D cone-beam CT (4DCBCT) images from sparse datasets. Our approach combines the idea of ‘bin-sharing’ with a deep convolutional neural network (CNN) model. More specifically, for each respiratory bin, an initial estimate of the patient sinogram is obtained by taking projections from adjacent bins and performing linear interpolation. Subsequently, the estimated sinogram is propagated through a CNN that predicts a full, high-quality sinogram. Lastly, the predicted sinogram is reconstructed with traditional CBCT algorithms such as the Feldkamp, Davis and Kress (FDK) method. The CNN model, which we referred to as the Sino-Net, was trained under different loss functions. We assessed the performance of the proposed method in terms of image quality metrics (mean square error, mean absolute error, peak signal-to-noise ratio and structural similarity) and tumor motion accuracy (tumor centroid deviation with respect to the ground truth). Overall, the presented prototype model was able to substantially improve the quality of 4DCBCT images, removing most of the streak artifacts and decreasing the noise with respect to the standard FDK reconstructions. The tumor centroid deviations with respect to the ground truth predicted by our method were approximately 0.5 mm, on average (maximum deviation was approximately 2 mm). These preliminary results are promising and encourage us to further investigate the performance of our model under more challenging imaging conditions and compare it against the state-of-the-art CBCT reconstruction algorithms. 

\end{abstract}

\keywords{4DCBCT, Deep Learning, Sparse Data, Image Reconstruction, Respiratory Motion
}

\section{INTRODUCTION AND RESEARCH GOAL}
\label{sec:intro} 

Cone beam computed tomography (CBCT) is a popular imaging modality in radiation therapy used for patient positioning, setup, and target localization directly preceeding the delivery of radiation. When scanning areas of the thorax or abdomen, respiration can create motion artifacts that are represented as blurring in the final reconstructed image. To alleviate this problem, 4DCBCT is used to temporally bin the projections acquired using a respiratory signal extracted from the patient's breathing, and reconstructing a volume for each phase\cite{Sonke}. Typical CBCTs are performed in 1 minute and scanned under low dose settings keeping the number of projections low resulting in 4DCBCTs images containing heavy streaking artifacts and of little use clinically. 

In this abstract, we propose for the first time a novel approach to reconstruct high-quality, geometrically accurate 4DCBCT images from sparse projection datasets and we assess its performance on a prototype dataset. Our method introduces the idea of sinogram ‘bin-sharing’, and combines it with a deep convolutional neural network (CNN) model to enhance the image quality and tumor location accuracy in 4DCBCT.  

\section{METHODS} 
\subsection{The bin-sharing method}
In order to reconstruct a 4DCBCT dataset using the bin-sharing method, the following steps were applied. Firstly, the patient respiratory signal was used to divide the measured projections (680 in total) into 10 bins, resulting in 55-85 projections per bin. Each subset of the projection data represent the so-called sparse sinograms (Figure 1 A). Second, each sparse sinogram was enhanced by taking the projections from two adjacent bins (i.e., the bin-sharing step). This approximation was made under the assumption that the patient respiratory motion between two consecutive bins is small. Third, the remaining missing projections in the sinogram of each bin were filled by linearly interpolated projections (Figure 1B). Subsequently, the shared+interpolated sinogram was propagated through a trained deep convolutional neural network (CNN) to enhance its quality (Figure 1C). Finally, each predicted high-quality sinogram was reconstructed using the standard Feldkamp, Davis and Kress CBCT algorithm \cite{Feldkamp:84}. The CNN was trained using pairs of input shared+interpolated sinogram images (as those shown in Figure 1B) and target full sinogram images (Figure 1D). The network model, training parameters and performance assessment are described in the following sub-sections.
\subsection{Patient Data}
Cone beam CT sinogram data used to train and validate the neural network, as well as to assess the overall performance of the proposed method, were generated from 4DCT datasets. Each 4D dataset consisted of 10 CT images of patients with lung tumors (respiratory phases) and their corresponding respiratory signal. More specifically, each CT image was forward projected under the CBCT geometry onto a set of 680 projections covering 360 degrees around the patient using the Reconstruction Toolkit Software \cite{rit2014}. Each projection image was 512 x 384 pixels in size. These complete sinograms (of size 680 x 512 x 384) were considered as the ground truth and the corresponding low-quality sinograms were created following the bin-sharing + interpolation method. In total, 4DCT datasets of 16 different patients were included in this work. These data, in addition to the respiratory signals and the CBCT scanning geometries, were facilitated by the organizers of the SPARE Challenge \cite{spare}. 

\begin{figure}[!htbp]
\centerline{\includegraphics[width=0.8\textwidth]{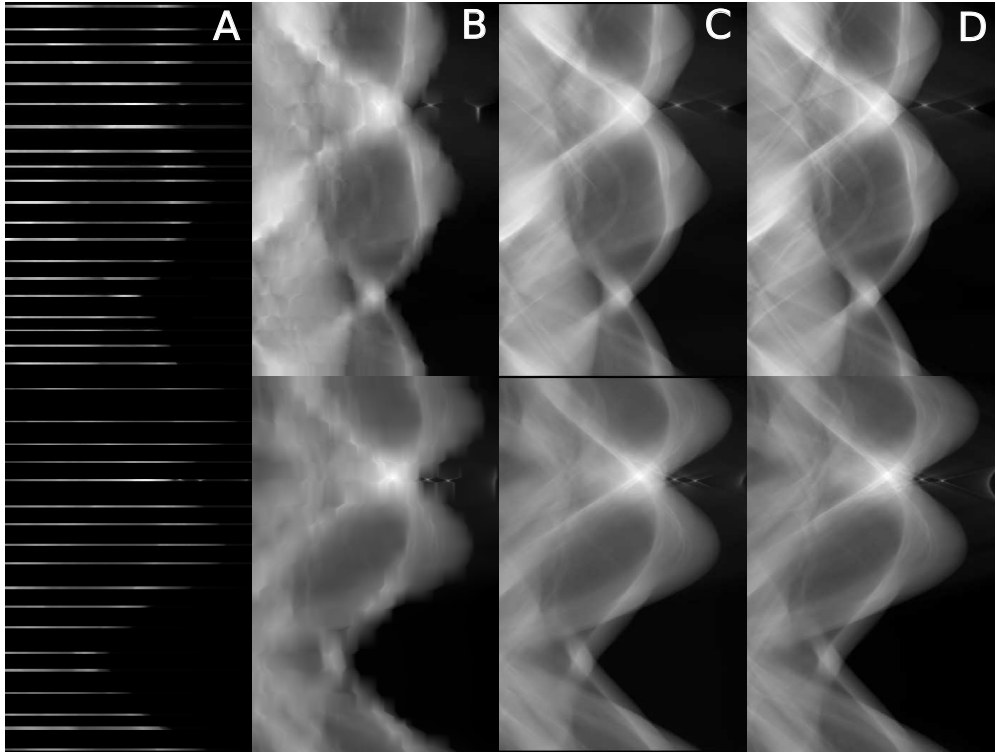}}
\caption{From left to right: sparse sinogram, shared+interpolation, sino-net output and ground truth of Patient 1 of the Test dataset.}
\label{fig_sino}
\end{figure}

\subsection{Network Architecture and Training}
\begin{figure}[tbp]
\centerline{\includegraphics[width=0.8\textwidth]{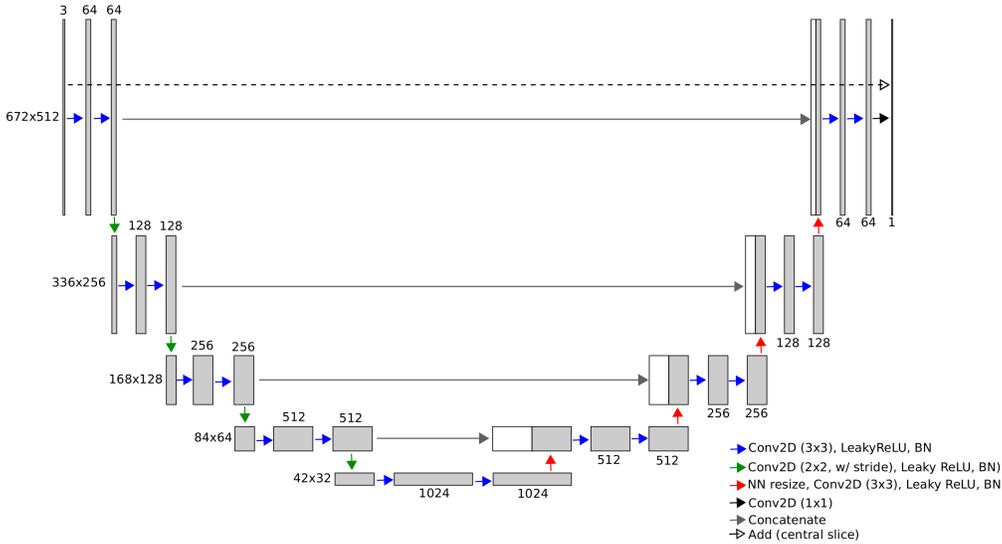}}
\caption{Sino-Net architecture.}
\label{fig_sino_arch}
\end{figure}

The architecture of our CNN (which we refer to as `Sino-Net') is based on the popular `U-Net' model proposed by Ronneberger et al 2015\cite{DBLP:journals/corr/RonnebergerFB15} but with minor modifications. In particular, we replaced the max-pooling and the de-convolution layers by 2D-convolutions to prevent checkerboard artifacts\cite{Odena}. Additionally, we added a residual connection between the inputs and the output to facilitate learning\cite{Li2017}. The details of Sino-Net architecture, including the number and size of feature maps, activations, and convolutional layers are shown in Fig.~\ref{fig_sino_arch}. The U-Net network architecture has shown success in sparse CT reconstruction\cite{Lee2018, Han2016} and restoration of Low-Dose CT images\cite{Li2017}. 

Following the philosophy of Zhao et al. 2015\cite{Zhao2015}, we explored the performance of the Sino-Net under three different loss-functions: the mean absolute error (or $L1$), the perceptually motivated multi-scale similarity index\cite{Wang2003} (MS-SSIM) and a weighted sum of $L1$ and MS-SSIM (with weights equal to 0.16 and 0.84, respectively, as determined by Zhao et al.). Each Sino-Net model was trained with sinogram data from 12 patients. In order to allow for remainderless downsampling of feature maps by a factor of 2, the initial image height was cropped from 680 down to 672. To minimize aliasing effects when predicted sinograms are stacked back together\cite{DBLP:journals/corr/abs-1710-03344}, each input image consisted of 3-consecutive 672 x 384 low-quality sinogram slices (Figure 1B), whereas the target output was the high-quality sinogram slice corresponding to the central slice of the input (Figure 1D). Each patient full sinogram was normalized to [0,1] and no additional preprocessing was applied. In total, 30720 sinogram slices were used for training corresponding to 512 slices x 12 patients x 5 bins. Note that in order to minimize computational cost, only 5 out of the 10 sinogram bins (namely bins \#1, \#3, \#5, \#7 and \#9) of each patient were used during training. The training data was augmented by randomly flipping the sinograms horizontally. 

The network model was implemented in Keras\cite{chollet2015keras}  (with Tensorflow backend\cite{tensorflow2015-whitepaper}) and trained using the Adam optimizer algorithm with an initial learning rate of $10^{-4}$. All training runs were carried out on GoogleCloud using a Nvidia Tesla P100 GPU. In this hardware, the time to run one epoch was approximately 2050 seconds. Due to the large size of the input images, the mini-batch size was set to 2 in order to fit into the GPU memory. The validation loss was monitored during training and when reaching a plateau the initial learning for the following epoch was decreased by a factor of 10. When the loss no longer decreased the training was halted, resulting in a total of 6 training epochs.

\subsection{Performance assessment}
We assessed the performance of the Sino-Net models (trained under each investigated loss function) in terms of the image quality and tumor geometric accuracy of 4DCBCT images reconstructed from the estimated high-quality sinograms on a validation and a test set. The validation and test sets contained two patient datasets each. The validation set was used to determine the best performing loss function which was subsequently applied to the test set. Since the location of the centroids of tumors present in the patient data were not available to us (as it required tumor segmentation performed by a physician), we placed artificial uniform tumors in the lungs of each patient (spheres with diameter = 3 cm). Additionally, to model tumor motion due to breathing, we applied a sinusoidal motion to each tumor in both the inferior-superior (2 cm in amplitude) and anterior-posterior (1 cm in amplitude) directions on each of the 4DCT images of the validation and test set.   

For image quality assessment, the mean-square error (MSE), mean-absolute error (MAE), peak signal-to-noise ratio (PSNR) and structural similarity index (SSIM) between the estimated and the ground truths 4DCBCTs were computed. The tumor geometric accuracy was evaluated in terms of the mean absolute deviation of the tumor centroid with respect to the ground truth (along the X, Y and Z axes in the reconstructed image) and the Sorensen-DICE coefficient. To determine the tumor centroid, tumors were automatically segmented by thresholding a 4 cm diameter sphere around the tumor region with a threshold such that the recovered tumor volume was equal to the true volume.

\section{RESULTS}
\subsection{Image Quality}
Table.~\ref{tab:geometry} presents the summary of the image quality assessment results for both the validation and test set. To allow for comparisons the MSE and MAE were normalized to the dynamic range of each patient image. The model using the loss function defined by Zhao et al. in  Ref.~\citenum{Zhao2015} had the best performance for three out of the four image-quality metrics evaluated and was deemed the most suitable for further reconstructions of the test set. Overall, the selected model generalized relatively well to the test dataset. 

\begin{figure}[!htbp]
\centerline{\includegraphics[scale=0.38]{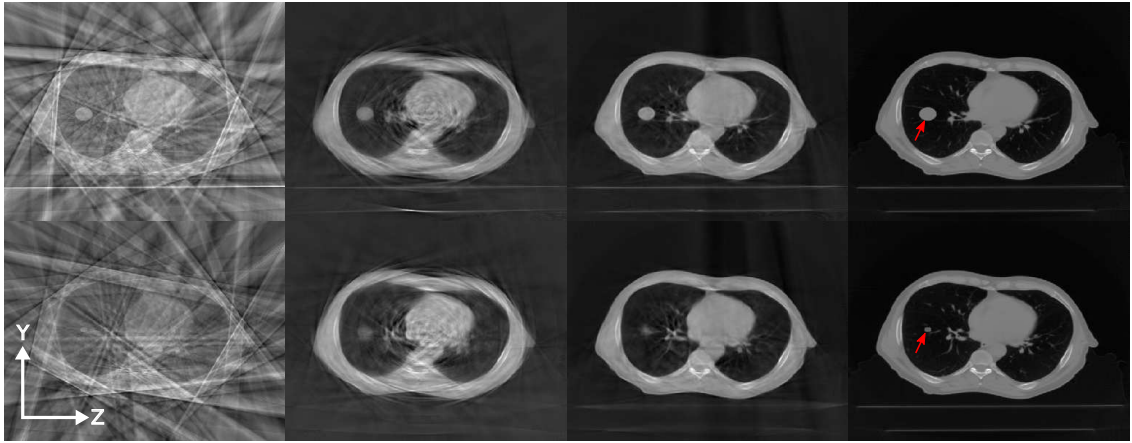}}
\caption{Axial slices from CBCT FDK reconstructions generated from sparse (far left), shared interpolation, Sino-Net prediction, and ground truth (far right) sinograms for Test Patient 1. Top row: CBCT reconstruction of Bin \#1, Bottom row:  Bin \#5. The red arrows indicate the artificial tumor.}
\label{fig_axial_sino}
\end{figure}

\begin{figure}[!htbp]
\centerline{\includegraphics[scale=0.42]{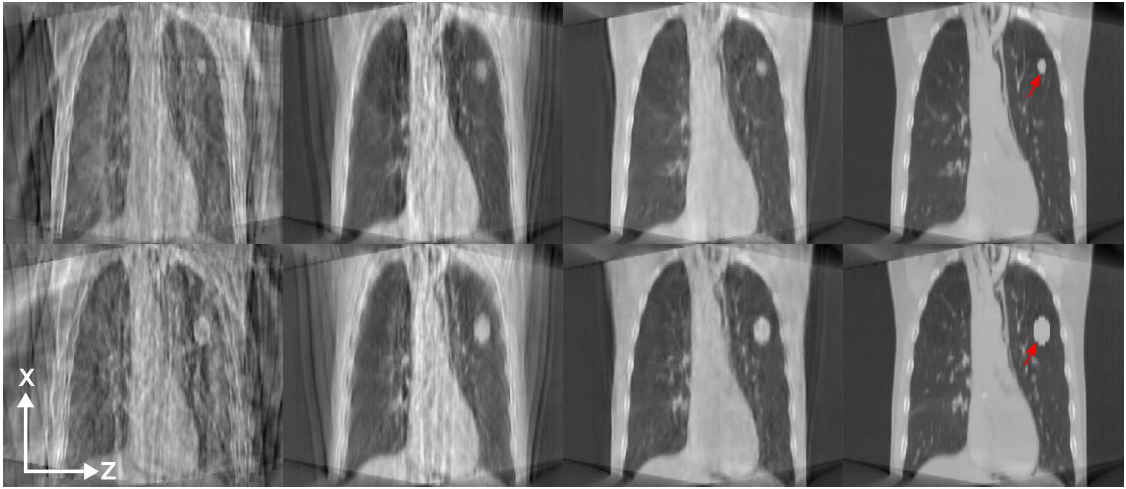}}
\caption{Coronal slices from CBCT FDK reconstructions generated from sparse (far left), shared interpolation, Sino-Net prediction, and ground truth (far right) sinograms for Test Patient 2. Top row: CBCT reconstruction of Bin \#1, Bottom row:  Bin \#5. The red arrows indicate the artificial tumor.}
\label{fig_coronal_sino}
\end{figure}

An example of the CBCTs reconstructed from the Sino-Net predictions on the test dataset, along with the sparse, the shared-bin+interpolation, and ground truth sinograms are shown in Fig.~\ref{fig_axial_sino} and Fig.~\ref{fig_coronal_sino}. It is worth noting the severity of the artifacts present in the CBCT images reconstructed from sparse sinograms, mostly due to the low number of projections available and the non-uniform distribution of these projections around the patient (see Fig.~\ref{fig_sino}A). Ring-like artifacts as well as severe blurring are also present in the share+interpolated reconstructions.

\subsection{Tumor motion and geometric accuracy}
\begin{figure}[!htbp]
\centerline{\includegraphics[scale=0.45]{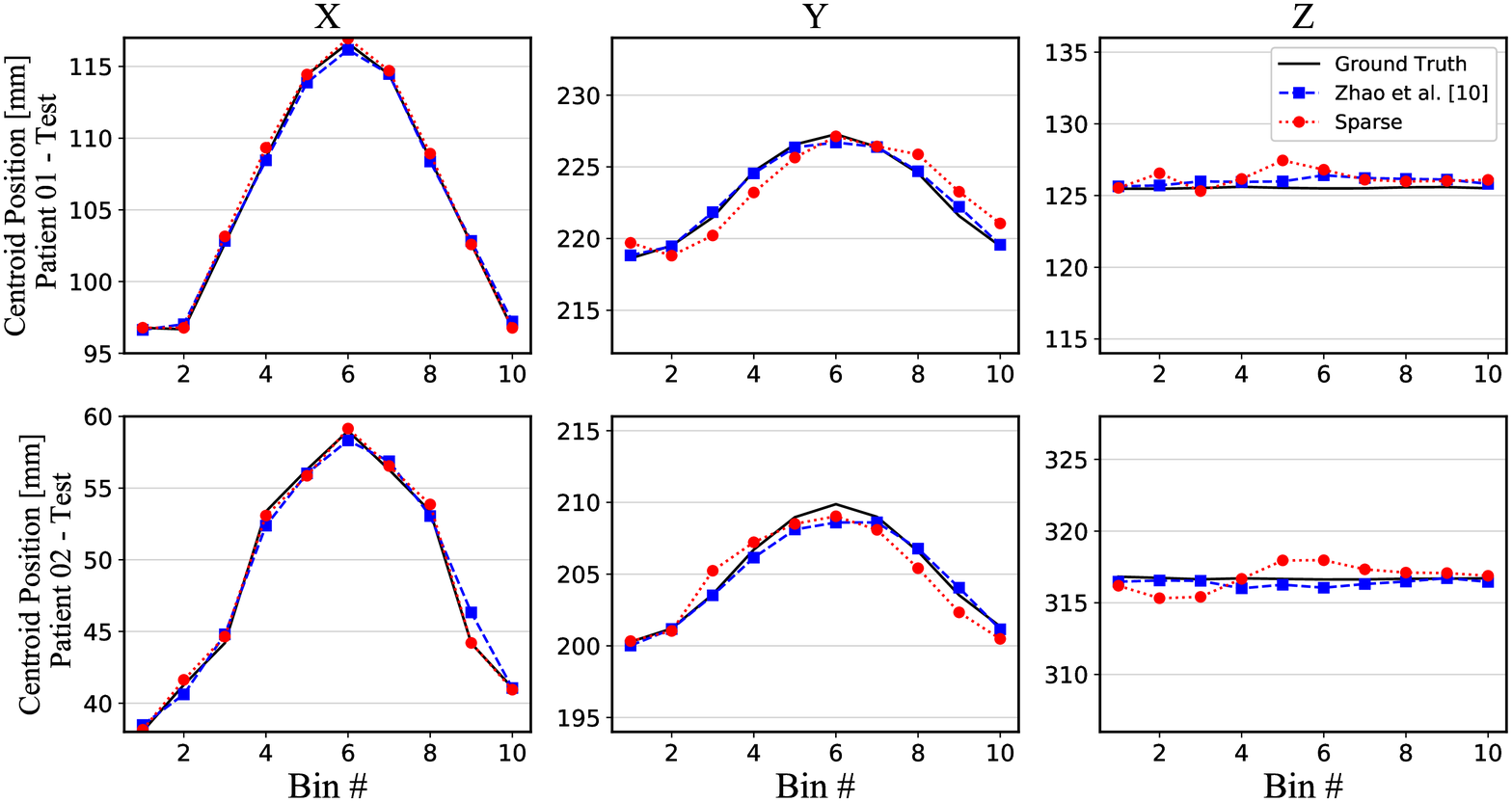}}
\caption{Tumor centroid position in different respiratory bins.}
\label{fig_tumor_motion}
\end{figure}

The mean absolute deviations of the tumor centroids and Dice coefficients are presented in Table~\ref{tab:geometry}. Similar to the image quality assessment, the best performing method was that of Zhao et al.~\cite{Zhao2015}. On average, tumor centroid deviations on reconstructed CBCTs were within 0.5 mm from the ground truth, with some reconstructed bins having deviations as large as 2 mm. Fig.~\ref{fig_tumor_motion} illustrates the centroid position as a function of respiratory bin for the two patients in the test dataset.

The current tumor motion results, although encouraging, represent the ideal case scenario where tumors are uniform, high contrast spheres. In fact, even the sparse reconstructions, despite of their low image quality, were effective at recovering the tumor location with high accuracy. In the future, tumors with irregular shapes and decreasing contrast will be added to the patient datasets to evaluate the methods under more realistic conditions. 

\begin{table}[tbp]
\caption{Image quality/Geometric accuracy assessment of the three models investigated in our study. The Mean-Squared-Error (MSE) and Mean-Absolute-Error (MAE) are normalized to the dynamic range of the tested patient images. Quantities in bracket represent the standard deviations of each evaluated metric across all patient images. Metrics in bold represent those that achieved the highest performance across our selected loss functions.} 
\label{tab:geometry}
\centering
\resizebox{1.0\columnwidth}{!}{%
\begin{tabular}{l|cccc|cccc|}
\cline{2-9}
   & \multicolumn{4}{c|}{Image Quality} & \multicolumn{4}{c|}{Geometric Accuracy} \\ \cline{2-9} 
   & MSE ($10^{-4}$) & MAE ($10^{-2}$) & PSNR [dB] & SSIM & \multicolumn{1}{c}{$\Delta x$[mm]} & \multicolumn{1}{c}{$\Delta y$[mm]} & \multicolumn{1}{c}{$\Delta z$[mm]} & \multicolumn{1}{c|}{DICE}  \\ \hline
   \multicolumn{1}{|c}{} & \multicolumn{8}{|c|}{Validation} \\ \hline
 \multicolumn{1}{|l|}{L1} &2.45 (0.40) & 1.15 (0.08)  & 36.2 (0.7) & 0.869 (0.015) & 0.6 (0.5) & 0.4 (0.3) & 0.3 (0.2) & 0.89 (0.04) \\
 \multicolumn{1}{|l|}{MS-SSIM} & 2.31 (0.37) & 1.11 (0.09) & 36.4 (0.6) & \textbf{0.911 (0.018)} & 0.6 (0.4) & 0.3 (0.2) & 0.3 (0.2) &  0.91 (0.03)\\
 \multicolumn{1}{|l|}{Zhao et al. \cite{Zhao2015}} & \textbf{2.25 (0.37)} & \textbf{1.09 (0.09)} & \textbf{36.5 (0.7)} & 0.873 (0.011) & 0.5 (0.4) & \textbf{0.3 (0.2)} & \textbf{0.2 (0.2)} & \textbf{0.91 (0.03)} \\ 
  \multicolumn{1}{|l|}{Sparse} & 311.5 (80.4) & 16.1 (2.8) & 15.2 (1.1) & 0.070 (0.026) & \textbf{0.3 (0.2)} & 0.7 (0.5) & 0.5 (0.4) & 0.85 (0.04)\\ \hline
     \multicolumn{1}{|c}{} & \multicolumn{8}{|c|}{Test} \\ \hline
 \multicolumn{1}{|l|}{Zhao et al. \cite{Zhao2015}} &  4.66 (1.58) & 1.55 (0.28) & 33.6 (1.5) & 0.849 (0.017) & 0.5 (0.5) & 0.3 (0.3) & 0.4 (0.2) & 0.91 (0.02) \\
 \multicolumn{1}{|l|}{Sparse} & 439 (191) & 17.2 (4.5) & 14.0 (2.0) & 0.063 (0.011) & 0.3 (0.2) & 0.9 (0.5) & 0.7 (0.5) & 0.84 (0.04)\\ \hline
\end{tabular}
}
\end{table}

\newpage
\section{CONCLUSIONS AND FUTURE WORK}

The proposed 4DCBCT reconstruction method, which combines sinogram bin-sharing and a deep learning approach, allows us to obtain high-quality and geometrically accurate 4DCBCT images reconstructed from sparse datasets. 

Despite the promising results presented in this abstract, there are several aspects that we plan to explore in the future in order to assess the robustness and performance of our method. In particular, we plan to evaluate tumor geometric accuracy under tumors with low contrast and irregular shapes. Additionally, we intend to assess the performance on a large dataset containing a variety of patient anatomies and sizes. Finally, we will compare the performance of our method against other neural networks architectures and state-of-the-art CBCT reconstruction algorithms.

\acknowledgments     
The authors thank the team behind the The SPARE challenge for inspiring this work and sharing the patient dataset. The SPARE team is led by Dr Andy Shieh and Prof Paul Keall at the ACRF Image X Institute, The University of Sydney. Collaborators who have contributed to the datasets include Prof Xun Jia, Miss Yesenia Gonzalez, and Mr Bin Li from the University of Texas Southwestern Medical Center, and Dr Simon Rit from the Creatis Medical Imaging Research Center.


\bibliography{JPNET_bib}   
\bibliographystyle{spiebib}   

\end{document}